\documentclass[12pt]{amsart}
\usepackage{amsmath, amssymb, amscd, array}

\usepackage{hyperref}

\usepackage[utf8]{inputenc}
\usepackage{amsmath}
\usepackage{amssymb}
\usepackage{csquotes}
\usepackage{mathtools}
\usepackage{dirtytalk}
\usepackage{fixltx2e}
\usepackage{DLdef1}

\begin{document}

\title{Non-split superstrings of dimension $(1|2)$}

\author{Dimitry Leites${}^{a,*}$, Alexander S. Tikhomirov${}^{b}$}
\address{${}^*$Corresponding author\\
${}^a$Department of Mathematics\\
Stockholm University, Albanov\"agen 28, SE-114 19 
Stockholm, Sweden\\
\email{dimleites@gmail.com}\\
${}^b$Faculty of Mathematics\\
National Research University, Higher School of Economics, Usacheva st. 6, 119 048 
Moscow, Russia\\
\email{astikhomirov@mail.ru}\\
}
%\fi

\makeatletter
\@namedef{subjclassname@2020}{\textup{2020} Mathematics Subject Classification}
\makeatother

\subjclass[2020]{Primary 58C50, 32C11, 81T30}

\keywords{Non-split supermanifold}

\begin{abstract} Any supermanifold diffeomorphic to one whose structure sheaf is the sheaf of sections of a~vector bundle over the underlying manifold is called split. Gaw\c{e}dzki (1977) and Batchelor (1979) were the first to prove that any smooth supermanifold is split. In 1981, P.~Green, and Palamodov, found examples of non-split analytic supermanifolds and described  obstructions to splitness that were further studied by Manin (resp. Onishchik with his students) following Palamodov's (resp. Green's) approach.  Following Palamodov, Donagi and Witten demonstrated that some of the moduli supervarieties of superstring theories are non-split. None of the above-mentioned authors considered  odd parameters of supervarieties of obstructions to non-splitness. Here, using Palamodov's approach, we classify and describe the even (degree-2) and the odd (degree-1) obstructions to splitness of $(1|2)$-dimensional superstrings. In particular, we correct calculations of degree-2 obstructions due to Bunegina and Onishchik and confirm Manin's answer. \end{abstract} 

\maketitle

\markboth{\itshape Dimitry Leites, Alexander S. Tikhomirov}
{{\itshape Non-split superstrings of dimension $(1|2)$}}

\thispagestyle{empty}

\section{Introduction} \label{Intro}
Let $M$ be a~real manifold, $\textbf{E}\tto M$ a~vector bundle with fiber $V_x$ over $x\in M$. Let $\cL_{E(\textbf{E})}$ be the sheaf of sections of the exterior algebra of the bundle $\textbf{E}$ whose fiber at $x\in M$ is the exterior algebra $E(V_x)$ of $V_x$. Any supermanifold diffeomorphic to one of the form $\cM=(M, \cL_{E(\textbf{E})})$ is called \textit{split}. In the category of smooth supermanifolds, every object is split; for a~transparent proof of this fact, see \cite[Subsection 4.1.3]{MaG}, which is more instructive than the first publications \cite{Ga, Bat}.

To any supermanifold
$\cM=(M, {\mathcal O}_\cM)$, where ${\mathcal O}_\cM$ is a~filtered sheaf of supercommutative superalgebras, there corresponds a~supermanifold with the graded sheaf constructed as follows. Consider the filtration
\begin{equation}
{\mathcal O}_\cM := {\mathcal J}^{0}\supset{\mathcal J}^{1} \supset {\mathcal J}^{2}
\supset \cdots\label{(1.12)}
\end{equation}
of ${\mathcal O}_\cM$ by the powers of the subsheaf of ideals ${\mathcal J}:={\mathcal J}^{1}$ generated by the odd elements that span the fiber of a~bundle $\textbf{E}$. The associated graded sheaf
$$
\gr{\mathcal O}_\cM=\bigoplus_{k\ge 0}\gr^k{\mathcal O}_\cM,
$$
where $\gr^{k}{\mathcal O}_\cM = {\mathcal J}^{k}/{\mathcal J}^{k+1}$,
gives rise to the supermanifold $(M,\gr{\mathcal O}_\cM)$, which is split (or is a~ singular version of a~split supervariety, if the stalk of $\gr{\mathcal O}_\cM$ is not a~free Grassmann algebra, but a~quotient thereof). The supermanifold $(M,\gr{\mathcal O}_\cM)$ is
called the \textit{retract} of~ $\cM$. 

In \cite{Gr}, Green showed that in the category of complex-analytic supermanifolds there exist non-split supermanifolds and gave the first such example.

While editing the manuscript of Berezin's posthumous book, Palamodov completely rewrote (see \cite[Ch.~4]{Ber1} and \cite[Ch.3, \S\S4--7]{Ber2}) the draft of the chapter devoted to non-split supermanifolds and also gave examples of non-split supermanifolds expounding his paper \cite{Pa}. Palamodov's description of obstructions to splitness looks totally different from Green's whom Palamodov does not cite in \cite{Pa} submitted on November 23, 1981, obviously being unaware of \cite{Gr} submitted on September 24, 1981.

Later on, Manin described non-split supermanifolds with the underlying manifold $\Cee\Pee^m$, in particular, he classified degree-2 obstructions to splitness of supermanifolds whose retract is the split superstring  $\cC\cP^{1|2}$, a~projective supermanifold. Manin's way of describing non-split supermanifolds is the same as Palamodov's, see \cite{Pa}, \cite[Ch.~4]{Ber1}, \cite[Ch.~3]{Ber2}; however, Manin did not mention either Palamodov or Green whose way of describing non-splitness is unclear to us, unlike Palamodov's. 

Manin's classification of supermanifolds whose retract is the projective superspace $\cC\cP^{1|2}$, a.k.a. superstring, differs from that given by Bunegina and Onishchik, who followed the method outlined by Green. Since the answers are different, at least one of them is wrong; our description of degree-2 invariants coincides with Manin's. We cannot pinpoint the exact erroneous spot in \cite{BO}; we used the same transparent approach of Palamodov as Manin. Donagi and Witten showed that some of the moduli supervarieties of superstring theories are non-split, see \cite{DW}.

None of the above-mentioned authors, and none of the authors of the collection \cite{LB} and references therein, considered odd parameters of supervarieties of obstructions to splitness.
To consider such obstructions, recall, see \cite{L}, that functorially, for any (say, finitely generated, over the same ground field) supercommutative superalgebra~ $C$, the changes of coordinates in a~chart of a~superdomain $\cU$ with even coordinates $u= (u_1, \dots, u_m)$ and odd coordinates $\xi=(\xi_1, \dots, \xi_n)$ are given by the parity-preserving $C$-linear automorphisms of the form 
\begin{equation}
\label{trFun}
 \left\{\renewcommand{\arraystretch}{1.6}
 \begin{array}{ll}
 \varphi^*(u_i) \;& =\; \varphi_i(u, \xi):= \varphi^0_i(u) +
 \fbox{$\sum\limits_{r\geq 1}\;
 \sum\limits_{i_1<\dots<i_{r}}\varphi_i^{i_1\dots
 i_{r}}(u)\xi_{i_{1}}\cdots \xi_{i_{r}}$} \text{~~for all $i$}, \\
\varphi^*(\xi_j) \;&=\; \psi_j(u, \xi):= \sum\limits_{r\geq 0}\; \sum\limits_{j_1<\dots <j_{2r+1}}
 \psi_j^{j_1\dots
 j_{2r+1}}(u)\xi_{j_{1}}\cdots\xi_{j_{2r+1}}\\
&+\fbox{$\psi^0_j(u)+
\sum\limits_{r\geq 1}\; \sum\limits_{j_1<\dots <j_{2r}}
 \psi_j^{j_1\dots
 j_{2r}}(u)\xi_{j_{1}}\dots\xi_{j_{2r}}$}\text{~~for all $j$},\\
\end{array}
\right.
\end{equation}
where, for all $r$, the even and the odd parameters are as follows
\[
\begin{array}{l}
\varphi^0_i(u),\ \ \varphi_i^{i_1\dots
 i_{2r}}(u), \ \ \psi_j^{j_1\dots
 j_{2r+1}}(u)\in C_\ev\ \ (\text{the even parameters}),\\
 \psi^0_j(u),\ \ \psi_j^{j_1\dots
 j_{2r}}(u),\ \ \varphi_i^{i_1\dots
 i_{2r+1}}(u)\in C_\od \ \ (\text{the odd parameters}).
\end{array}
 \]
These are parameters 
of the infinite-dimensional supergroup of automorphisms of $C^\infty(\cU)$ or, equivalently, of diffeomorphisms of $\cU$; infinitesimally: of the Lie superalgebra $\fvect(m|n)$.

Note that the boxed summands in $ \varphi^*(u_i)$ are precisely what the prefix ``super" brings to the non-super (``classical") Differential Geometry: these summands, meaningless in the non-super setting even for the even parameters $\varphi_i^{i_1\dots
 i_{2r}}(u)$, acquire meaning, as explained above, even for the odd parameters $\varphi_i^{i_1\dots i_{2r+1}}(u)$. The split supermanifolds are those for which there exists an atlas without these boxed summands in $ \varphi^*(u_i)$.

\section{Deformations of superstrings ${\mathcal{CP}}^{1|m}$ for $m=1$ and $2$}

For descriptions of non-split superstrings corresponding to bundles of rank $m\leq 3$ over ${\mathbb{CP}}^1$, see \cite{BO, MaG}. Our results below correct \textit{these} answers in \cite{BO} and in \cite[Ch.4, \S2, Subsection~10(a), p. 192]{MaG} in the cases $m=1$ and~$m=2$. Our arguments  clearly show that the cases of $m>2$ are much more complicated than claimed in \cite[Theorem 1 ($m=3$)]{BO}. 

\ssec{Basics: recapitulation} Let us cover ${\mathbb{CP}}^1$ by two affine charts $U_0$ and $U_1$, with local coordinates
$x$ and $y=x^{-1}$, respectively. Let $\xi$ (resp., $\eta$) be a~ basis section of the line bundle ${\mathbf L}_{k}$ over~
$U_0$ (resp., $U_1$) such that
the transition functions 
in $U_0\cap U_1$ are of the form
\[ 
y ={x}^{-1},\ \ \ \
\eta ={x}^{k}\xi\quad \text{(up to a~constant factor). }
\]
This bundle ${\mathbf L}_{k}$ is said to be \textit{of degree} $k\in\Zee$. The sheaf of sections of the bundle ${\mathbf L}_{k}$ is denoted $\cO(k)$. In particular, $\cO:=\cO(0)$ is the structure sheaf (of functions), and since $dy=-x^{-2}dx$, then the cotangent sheaf of volume forms $\Omega^1$ is $\cO(-2)$, and hence ($\partial_y=x^{2}\partial_x$) the dual to~ $\Omega^1$ tangent sheaf is $\cO(2)$.

A~theorem of Grothendieck  (see \cite{HM}) states that any holomorphic bundle $\textbf{E}$
over ${\mathbb{CP}}^1$ is a~direct sum of line bundles ${\mathbf L}_{-k_i}$ with uniquely determined degrees $k_1, k_2,\dots$ 
In the interesting case  we are going to consider where the supermanifold $\cM=({\mathbb{CP}}^1, \cL_{E(\textbf{E})})$ is homogeneous, 
the numbers $k_i$ must be non-negative, see \cite[Proposition 13]{BO}, i.e., hereafter
\[
\text{${\mathbf{E}}=
{\mathbf{L}}_{-k_1}\oplus{\mathbf{L}}_{-k_2}\oplus{\mathbf{L}}_{-k_3}
\oplus{\mathbf{L}}_{-k_4}\oplus\dots$,\quad where
$k_1\geq k_2\geq k_3\geq k_4\geq \dots \geq 0$}.
\]

We will need facts \eqref{facts1}--\eqref{facts3}, see, e.g., \cite[\S~1.1]{OSS}, and the Serre dualization $\vee$, see \cite{H}:
\be\label{facts1}
%\begin{array}{l}
\cO(a)\otimes\cO(b)\simeq \cO(a+b);
\ee
\be\label{facts2}
\begin{array}{l}
\dim H^0(\Cee\Pee^{1}; \cO(a))=\begin{cases}a+1,&\text{if $a\geq0$},\\
0,&\text{otherwise};\end{cases}\\
\text{for a~basis of the space of sections $H^0_a$ over $U_0$ we can take $\xi, x\xi,\dots, x^{a}\xi$};\\
H^0_a\simeq\{P\in\Cee[x, y]\mid\deg P=a\};\\
\end{array}
\ee
\be\label{facts3}
\begin{array}{l}
\dim H^1(\Cee\Pee^{1}; \cO(a))=\begin{cases}-a-1,&\text{if $a\leq -2$},\\
0,&\text{otherwise};\end{cases}\\
\text{for a~basis of the space of sections $H^1_a$ over $U_0$ we can take}\\
\xi, x^{-1}\xi,\dots, x^{2+a}\xi \ \ \text{if $a\leq -2$};\\
\end{array}
\ee
\be\label{Serre}
H^1(\Cee\Pee^1; \cO(a))\simeq H^0(\Cee\Pee^1; \cO(-a-2))^\vee.
\ee

Palamodov showed (\cite{Pa}, \cite[Ch.~4]{Ber1}, \cite[Ch.~3]{Ber2}) that the obstructions to splitness of a~supermanifolds $\cM^{1|m}$ with retract $(\Cee\Pee^{1}; \cE^{\bcdot}(\textbf{E}))$, where $\rk \textbf{E}=m$, form the set of 
\be\label{i}
\text{$\Aut(\textbf{E})$-orbits in $\bigoplus_{i\geq 1}\ H^1(\Cee\Pee^{1}; \cO(2)\otimes \cE^{i}(\textbf{E}))$,}
\ee
where $\cE^{i}(\textbf{E})$ is the sheaf of sections of the $i$th exterior power $E^{i}(\textbf{E})$
of the bundle $\textbf{E}$. For no reason, all researchers considered, so far, only \textbf{even} values of $i$, thus ignoring odd parameters of supervarieties of obstructions to non-splitness.  

Recall that the set of points of the Grassmannian $\text{Gr}(k ,n)$ of $k$-dimensional subspaces in the $n$-dimensional vector space can be identified with the set of points of  the super Grassmannian $\text{SGr}(k ,n)$ of $(0|k)$-dimensional vector subsuperspaces in the $(0|n)$-dimensional superspace; the tautological bundles of  these Grassmannians are, however, different: the fiber over the point corresponding to the  $k$-dimensional subspace $V$ (resp.,  $(0|k)$-dimensional subspace $\Pi(V)$, where $\Pi$ is the inversion of parity functor) is $V$ (resp., $\Pi(V)$) itself, see \cite{MaG}.

\ssec{The simplest examples}

\underline{$m=1$} (from \cite{L}). By eq.~\eqref{facts3}, the only possible value of $i$ in eq.~\eqref{i} is $i=1$, so the odd obstructions to splitness of supermanifolds $\cM$ of complex superdimension $(1|1)$, a.k.a. superstrings, with underlying $\Cee\Pee^{1}$, whose retract corresponds to the line bundle $\mathbf{E}:= \mathbf{L}_{-k}$ with rank-1 fiber $\Cee$ are described by the \textit{non-zero}, i.e., different from the origin,  orbits under the action of $\GL(\Cee)\simeq\GL(1)$, which is also the group of automorphisms of $\mathbf E$ and hence acts in the purely odd space 
\be\label{Ha}
H_k:=H^1(\Cee\Pee^{1}; \cO(2)\otimes \cE^{1}(\textbf{E}))\simeq H^1(\Cee\Pee^{1}; \Pi(\cO(2-k));
\ee
i.e., the set of orbits is parameterized by $\Cee\Pee^{k-4}$ for $k\geq 4$, see eq.~\eqref{facts3}, whereas  for $k<4$ there are no obstructions to splitness of $\cM$. (This coincides with the result of \cite{L}, where notation ${\mathbf E:= \mathbf L_{k}}$ was used.)

\underline{$m=2$}. Here, we correct the description of $\Aut(\textbf{E})$ in \cite[Example 10 in \S2, Ch.4]{MaG}. Let ${\mathbf{E}}:=
{\mathbf{L}}_{-a}\oplus{\mathbf{L}}_{-b}$, where
$a\leq b$. We have to consider the two cases: (A) where $a=b$ and (B) where $a<b$, each case having two subcases: $i=1$ and $i=2$ in eq.~\eqref{i}. Over every point, the group of automorphisms of the fiber is isomorphic to
\[
\begin{cases}\GL(2)&\text{if $a=b$},\\
\text{B}=\left\{\begin{pmatrix}* & 0\\
*&*\end{pmatrix}\right\}\subset \GL(2)&\text{if $a<b$}.\end{cases}
\]

Set $H:=H^1(\Cee\Pee^{1};\ \bigoplus_i\, \cO(2)\otimes \cE^{i}(\cO(-a)\oplus \cO(-b)))$.

If $a=b$, the group of automorphisms of $\mathbf{E}=V\otimes {\mathbf{L}}_{-a}$ for an abstract 2-dimensional space $V$ is isomorphic to $\GL(V)=\GL(2)$, and hence acts on $H$. %We will say that the orbit is \textit{non-zero} if it is distinct from $\{0\}$.

If $a<b$, set $d:=b-a$.  Tensoring $\cO(-a)\oplus \cO(-b)$ by $\cO(a)$ we show that
\[
\End(\cO(-a)\oplus \cO(-b))\simeq\End(\cO\oplus \cO(-d)). 
\]
The group $\text{B}$ of automorphisms of the fiber does not act on $H$, instead we consider in $H$ the orbits of the group $\text{G}:=\Aut(\cO\oplus \cO(-d))$ of global automorphisms of $\mathbf{E}$. 

\textbf{The degree-1 obstructions to splitness} are the non-zero $\GL(V)$-orbits in the purely odd space $V\otimes H_a$, recall eq.~\eqref{Ha}, if $a=b$, or  the $\text{G}$-orbits if $a<b$ in
\be\label{i=1}
\begin{array}{l}
H^1(\Cee\Pee^{1}; \cO(2)\otimes \cE^{1}(\cO(-a)\oplus \cO(-b)))\simeq
H_a\oplus H_b.\\
%\text{where $H_a:= H^1(\Cee\Pee^{1}; \Pi(\cO(2-a)))$}.\\
\end{array}
\ee
%Let ${\mathbf{E}}:={\mathbf{L}}_{-a}\oplus{\mathbf{L}}_{-b}$ for $a<b$; let $d:=b-a$. 
Then, the elements of the group
\be\label{G}
\begin{array}{lll}\text{G}&:=&\Aut(\cO(2-a)\oplus \cO(2-b))\simeq\Aut(\cO(-d)\oplus \cO)\\
&=&\Aut(\cO(-d))\oplus\Hom( \cO(-d), \cO)\oplus\Aut(\cO) \simeq\Cee^\times\oplus\Cee^{d+1}\oplus\Cee^\times. \end{array}
\ee 
can be represented by matrices of the form
\[
\begin{pmatrix}
\Lambda &g\\
0&\mu\end{pmatrix}\ \ \text{where $\Lambda=\lambda \One_{d+1}$, \ $\lambda, \mu\in\Cee^\times$ and $g\in H^0(\Cee\Pee^1, \cO(d))\simeq \Cee^{d+1}$.}
\]

If $f_{a}\neq 0$, set 
\be\label{Sigma}
\Sigma:=\{(f_a, gf_a)\in (H_a\oplus H_b)\simeq (\Cee^{a-3}\oplus \Cee^{b-3})\mid f_a\in \Cee^{a-3}, \ g\in \Cee^{d+1}\}.
\ee

For $a=4$, we have $\Sigma=\Cee^{a-3}\oplus \Cee^{b-3}$ and $\Sigma/\text{G}=\{\bcdot\}$, see the proof of Theorem~\ref{a=b}.

For $a>4$, let  $\overline \Sigma$  be the complement to $\Sigma$, an open dense subset of $\Cee^{a-3}\oplus \Cee^{b-3}$.

\textbf{The degree-2 obstructions to splitness} are the non-zero $\GL(2)$-orbits if $a=b$ or $\text{G}$-orbits if $a<b$ in the space 
\be\label{i=2}
\begin{array}{l}
H^1(\Cee\Pee^{1}; \cO(2)\otimes \cE^{2}(\cO(-a)\oplus \cO(-b)))\simeq
H^1(\Cee\Pee^{1}; \cO(2-a-b)).\\
\end{array}
\ee

%\textbf{Particular cases}.

\sssbegin{Theorem}\label{a=b} \textup{(A) Let ${\mathbf{E}}=
{\mathbf{L}}_{-a}\oplus{\mathbf{L}}_{-a}=V\otimes{\mathbf{L}}_{-a}$}, see eq.\eqref{i=1}. 
Then, the obstructions to splitness of ${\mathcal{M}}$ with retract $(\Cee\Pee^{1}, \cE^{\bcdot}(\textbf{E}))$ are as follows.

\textup{(A1)} The degree-$1$ obstructions form the non-zero $\GL(V)=\GL(2)$-orbits in $V\otimes H_a$, namely 

if $a< 4$, then there are no odd obstructions;

if $a=4$, then the non-zero $\GL(V)$-orbit in $V\otimes H_a\simeq \Pi(V)$ is $\Pi(V)\setminus\{0\}$;

if $a>4$, then the rank-$2$ non-zero $\GL(V)$-orbits in $V\otimes H_a$ form $\coprod_{V'\in\text{SGr}(2,H_a)} (V\otimes V')^o$, where ${(V\otimes V')^o}$ is the set of indecomposable tensors in $V\otimes V'$ and where $\text{SGr}(2,H_a)$ is the Grassmannian of $(0|2)$-dimensional subspaces in $H_a$; the rank-$1$ tensors in $V\otimes H_a$ form the set of decomposable tensors, i.e., the cone  without $\{0\}$ over the $\Pee(V)\times \Pee(H_a)$; the non-zero $\GL(V)$-orbits in this cone form $\{\bcdot\}\times \Pee(H_a)\simeq \Pee(H_a)$.

\textup{(A2)} The degree-$2$ obstructions to splitness are the non-zero $\GL(V)$-orbits in $H^1(\Cee\Pee^{1}; \cO(2-2a))$, i.e., 
\[
\begin{cases}\Cee\Pee^{2a-4}, 
&\text{if $a\geq 2$,}\\
\text{none},&\text{otherwise}.\end{cases}
\]

\textup{(B) Let ${\mathbf{E}}=
{\mathbf{L}}_{-a}\oplus{\mathbf{L}}_{-b}$ for $a<b$}; let $d:=b-a$. The $\text{G}$-orbits in the spaces of obstructions are as follows.

\textup{(B1)} The degree-$1$ obstructions to splitness form the non-zero $\text{G}$-orbits in $H_a\oplus H_b$, namely 
\[
\Pee(H_b) %= \Pee(H^1(\Cee\Pee^1; \cO(2-b)))\simeq \Pee((H^0(\Cee\Pee^1; \cO(b-4)))^\vee)
\ \ \text{ if $H_a=0$,}
\]
whereas if $H_a\neq 0$, then the obstructions constitute a~ set of of dimension $2a-9$ for $a>4$  consisting of non-zero $\text{G}$-orbits in the set 
$\overline \Sigma:=(H_a\oplus H_b)\setminus
(\{0\}\oplus H_b)$, 
whereas  $\Sigma/\text{G}=\{\bcdot\}$ for $a=4$.

\textup{(B2)} The degree-$2$ obstructions are the non-zero $\text{G}$-orbits in $H^1(\Cee\Pee^{1}; \cO(2-a-b))$, i.e.,
\[
\begin{cases}\Cee\Pee^{a+b-4} 
&\text{if $a+b\geq 4$,}\\
\text{none},&\text{otherwise}.\end{cases}
\]
\end{Theorem}

\begin{proof} Claims (A2) and (B2) are clear from eq.~\eqref{facts3}.

(A1) Then, $\dim H_a=a-3$. We consider the following four subcases (a)--(d). 

(a) $\dim H_a=0\Longleftrightarrow a<4$. Then, there are no odd obstructions to splitness of $\cM$.

(b) $\dim H_a=1\Longleftrightarrow a=4$. Then, the non-zero $\GL(V)$-orbit in $V\otimes H_a\simeq \Pi(V)$ is $\Pi(V)\setminus\{0\}$.

(c) $\dim H_a=2\Longleftrightarrow a=5$. Then, let 
\[
\begin{array}{l}
\text{$V:= \langle v_1, v_2\rangle =\{v=y_1v_1+y_2v_2$ for any $y_i\in\Cee$\},}\\
\text{$ H_a:= \langle u_1, u_2\rangle =\{u=x_1u_1+x_2u_2$ for any $x_i\in\Cee$\}.}\\
\end{array}
\]
 Let $W:=V\otimes H_a$; let $K$ be the cone of decomposable tensors $w=v\otimes u$ for any $u\in H_a$ and $v\in V$.
Observe that $K$ is the closure of the image of the map $f:(\Cee^\times)^4 \to W$ given by the formula
\[
f: (\alpha_1, \alpha_2, \beta_1, \beta_2)\mapsto(w_{11}:=\alpha_1\beta_1, w_{12}:= \alpha_1\beta_2, w_{21}:=\alpha_2\beta_1, w_{22}:=\alpha_2\beta_2).
\]
The fiber of $f$ is isomorphic to $\Cee^\times$, since $f(\alpha_1, \alpha_2, \beta_1, \beta_2)=f(\alpha_1', \alpha_2', \beta_1', \beta_2')$ implies that there exists a~$\lambda\in\Cee^\times$ such that $\alpha_i'=\lambda \alpha_i$ and $\beta_i'=\lambda^{-1} \beta_i$.

Thus, $K$ is a~hypersurface in $W$; clearly, it is singled out by the equation
\be\label{eq*}
w_{11}w_{22}-w_{12}w_{21}=0.
\ee

For every non-zero $u\in H_a$, the 2-dimensional subspaces 
\[
V_{\langle u\rangle}:=\{v\otimes u\in W\mid u\in H_a\}\simeq V
\]
are $\GL(V)$-orbits in $W$. Therefore, 
\be\label{K*}
K^*:=K\setminus \{0\}=\coprod_{\langle u\rangle\in \Pee( H_a)} V_{\langle u\rangle}.
\ee

\sssbegin{Lemma} The set $W\setminus K$ is a~homogeneous space under the action of $\GL(V)$. \end{Lemma} 

\begin{proof}[Proof \nopoint] of the lemma. Since $\dim(W\setminus K)=\dim \GL(V)=4$, to describe the stationary subgroup of a~ point it suffices to prove that the stabilizer of a~vector $w\in W\setminus K$ is
\be\label{(1)}
\text{St}_{\GL(V)}w=\{\id\}~~\text{for any $w\in W\setminus K$}. 
\ee
Having selected a~basis of $V$ we get an isomorphism 
\[
\GL(V)\tto \GL(2), \ \ g\mapsto A_g.
\]
Accordingly, fixing the basis $\{e_{ij}:=v_i\otimes u_j\mid 1\leq i,j\leq 2\}$ of $W$ we get a~representation 
\be\label{2}
\GL(V)\tto \GL(4)\simeq \GL(W), \ \ g\mapsto \widetilde A_g:=\begin{pmatrix} A_g&0\\
0&A_g\end{pmatrix} .
\ee
Let now $w:=(w_{11}, \dots, w_{22})^t$ and $g\in\text{Stab}_{\GL(V)}w$. Then, by \eqref{2}, we have
\[
\widetilde A_g w=w \Longleftrightarrow A_g (w_{11}, w_{12})^t=(w_{11}, w_{12})^t\ \ \text{and}\ \ A_g (w_{21}, w_{22})^t=(w_{21}, w_{22})^t,
\]
hence
\[w_1:=(w_{11}, w_{12})^t\ \ \text{and}\ \ w_2:=(w_{21}, w_{22})^t\ \ \text{belong to $\Ker (A_g-1_2)$,}
\]
where $1_2$ is the unit matrix. If $\dim\Ker (A_g-1_2)=1$, then the vectors $w_1$ and $w_2$ are collinear, so their coordinates satisfy equation \eqref{eq*}, i.e., $w\in K$ contrary to our assumption that $w\in W\setminus K$. Hence, $\dim\Ker (A_g-1_2)=2$, i.e., $A_g=1_2$, and so equality \eqref{(1)} is proved.
\end{proof} 

(d) $\dim H_a>2\Longleftrightarrow a>5$. Clearly, the ranks of decomposable tensors $w$ can be equal to either 2 or 1. If the rank is equal to 2, then the non-zero $\GL(V)$-orbits in $V\otimes H_a$ are $\coprod_{V'\in\text{SGr}(2, H_a)} (V\otimes V')^o$, where $(V\otimes V')^o$ is the set of indecomposable tensors in $V\otimes V'$, see case~ (c). 

If the rank is equal to 1, then the set of decomposable tensors in $V\otimes H_a$ is the cone over $K^*$, see eq.~\eqref{K*}; the non-zero $\GL(V)$-orbits in this cone form $\{\bcdot\}\times \Pee(H_a)\simeq \Pee(H_a)$. 

(B1) If $d:=b-a>0$, then $\Hom(\cO, \cO(-d))=0$ and
\[
\End(\cO(-d)\oplus \cO)\simeq\End(\cO(-d))\oplus\Hom( \cO(-d), \cO)\oplus\End(\cO)\simeq\Cee \oplus\Cee^{d+1}\oplus\Cee .
\]

The $\text{G}$-action on $\mathcal{E}nd(\cO\oplus \cO(-d))$ is as follows:
\[
\begin{array}{l}
\text{G}\times H_a\oplus H_b \tto H_a\oplus H_b,\\ 
\text{G}\times (\Cee^{a-3}\oplus \Cee^{b-3})\tto \Cee^{a-3}\oplus \Cee^{b-3}.\\
\end{array}
\]
In other words, for $b>a\geq 4$, 
we have
\[
\begin{array}{l}
(\Cee^\times\oplus\Cee^{d+1}\oplus\Cee^\times)\times (\Cee^{a-3}\oplus \Cee^{b-3})\tto \Cee^{a-3}\oplus \Cee^{b-3},\\
(\lambda, g, \mu)\times (f_{a}, f_{b})\mapsto (\lambda f_{a}, \, gf_{a}+\mu f_{b}).\\
\end{array}
\]

Consider the two cases of $\text{G}$-action.

(I) $f_{a}=0$. Then, $(0, f_{b})\mapsto (0, \mu f_{b})$ for any $\mu \in\Cee^\times$, so the $\text{G}$-orbits in $\{0\}\oplus H_b$  
form a~set isomorphic to $\Pee(H_b)$.

(II) $f_{a}\neq 0$. Then, the stabilizer of the point under the $\text{G}$-action is (clearly, $\lambda=1$)
\[
\begin{array}{l}
\St_{\text{G}}(f_a, f_{b})= 
\{(1, g, \mu)\in \text{G}\mid (f_{a}, f_{b})=(f_{a}, gf_{a}+\mu f_{b})\}\\ 
\simeq \{(g, \mu)\in \Cee^{d+1}\times \Cee^\times \mid (1-\mu)f_{b}=gf_{a} \}. 
\end{array}
\]

Let $(f_a, f_{b})\in\Sigma$. Then the relation $(f_{a}, f_{b})=(f_{a}, gf_{a}+\mu f_{b})$ yields
$\St_{\text{G}}(f_a, f_{b})\simeq \Cee^\times$. Clearly, $\dim\Sigma=b-2$, see eq.~\eqref{Sigma}, hence (see eq.~\eqref{G})
\[
\dim (\Sigma/\text{G})= \dim\Sigma-(\dim \text{G} -1)=b-2-(d+2)=a-4.
\]
In particular, for $a=4$, we have $\Sigma=\Cee^{a-3}\oplus \Cee^{b-3}$ and $\Sigma/\text{G}=\{\bcdot\}$.

For $a>4$ and $(f_a, f_b)\in \overline \Sigma$, see eq.~\eqref{Sigma}, the relation $(1-\mu)f_{b}=gf_{a}$ implies $\mu=1$ and $g=0$, so that $\St_{\text{G}}(f_a, f_{b})=\id$. Therefore, 
\[
\dim (\overline \Sigma/\text{G})=a+b-6-(d+3)=2a-9.
\]

To describe the set $\overline \Sigma/\text{G}$ explicitly is an interesting \textbf{open question} for experts in algebraic geometry; its study is beyond the scope of this note.
\end{proof}

%\subsection{Remark} Observe that the orbit $\Pi(V)\setminus \{0\}$ in the purely odd space $\Pi(V)$

\subsection*{Disclosures} No conflict of interest. Data availability: 
The data used to support the findings of this study are included within the article.

%%%%%%%%%%%%%%%%%%%%%%%%%%%%%%%%%%%%%

\end{document}